\documentclass{iau-JDSS}
\usepackage{graphicx}
\title[Ultramassive stars] 
{Evolution of Ultramassive Stars}

\author[L. Yungelson]   
{Lev Yungelson
\affiliation{Institute of Astronomy, Pyatnitskaya 48, 119017 Moscow, Russia.\break email: lry@inasan.ru}}

\pubyear{2006}
\volume{Volume 14}  
\pagerange{119--126}
\date{?? and in revised form ??}
\jname{Highlights of Astronomy, Volume 14}
\editors{K.A. van der Hucht, ed.}
\begin{document}

\maketitle

\begin{abstract}
We show that even most massive initial solar composition stars hardly form
black holes with mass exceeding $\simeq 100\,M_\odot$.

\end{abstract}

We have calculated evolution of 60 to 1001\,$M_\odot$ nonrotating
solar metallicity 
stars from ZAMS 
through core helium-burning stage. $1001\,M_\odot$ is the upper mass limit
for ZAMS 
of solar metallicity  stars obtained by us (in agreement with \cite{ishii99}). 
In the absence of empirical data and the theory of mass-loss for very massive 
stars, we assumed that $\Gamma$-instability is responsible for mass loss and 
applied an \textit{ad hoc} mass-loss law 
$\dot{M}=\frac{L}{v_\infty c}\,\frac{1}{(1-L/L_{Edd})^{(\alpha-1/2)}}$.  
Here $L_{Edd}$ is Eddington luminosity in the outermost meshpoint of the model. 
Parameter $\alpha=0.25$\ was chosen such that the stars with mass 
60 to 120\,$M_\odot$ spend  $\lesssim 1$\% of their lifetime to the right of 
the Humphreys-Davidson limit in the HR-diagram. Computations were carried out by appropriately modified TWIN-version of
  \cite{eggleton71} code. 
Results are relevant to the evolution of the most massive stars  in the  Galactic
center  clusters or (still hypothetical) descendants of runaway stellar 
collisions in young dense clusters that are thought to be progenitors of 
black-hole accretors in some ultraluminous X-ray sources 
(see, e. g. \cite{spz_etal04}).   

Mass-loss rates both for H- and He-rich stars that follow from above given expression for $\dot{M}$\ are consistent with the limits implied for line-driven winds and with  $\dot{M}$\ found for the most massive stars, e. g., in Arches, Quintuplet, HeI-clusters in the Galactic center. 

Stars under consideration form oxygen-neon cores with minor admixture of carbon and have thin helium-carbon envelopes ($Y_s\approx 0.2-0.4$).  Initial-final mass relation obtained in calculations and the fate of stellar  remnants (as roughly inferred from the model calculations for helium stars, e. g. \cite{heger_woosley_he02}) are shown in the Figure. These results suggest that the products of runaway stellar collisions hardly can produce black holes with mass $\gtrsim 100\,M_\odot$.
\begin{center}
\begin{figure}[!h]
\hspace{3cm}
 \includegraphics[scale=0.4,angle=-90]{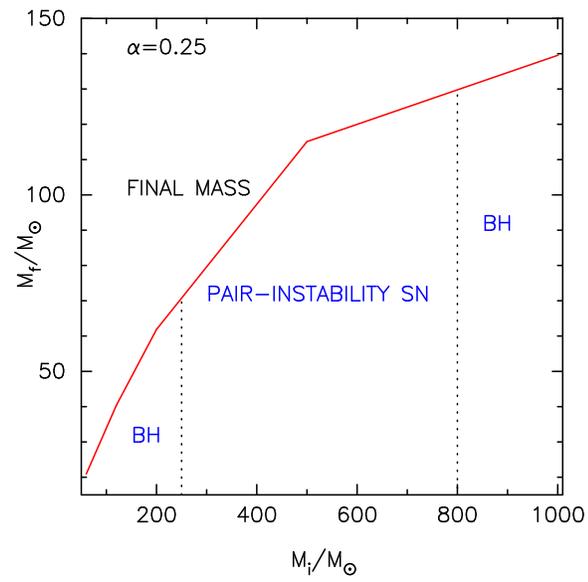}
  \caption{Relation between initial and final masses of stars and the nature of stellar remnants.}
\label{fig:wave}
\end{figure}
\end{center}
%

\begin{acknowledgments}
The author acknowledges P.P. Eggleton for providing his evolutionary code and support by IAU travel grant.
\end{acknowledgments}

\end{document}